\documentclass[aip,jmp,amsmath,amssymb,preprint,%linenumbers%reprint,%%author-year,%%author-numerical,%
]{revtex4-1}

\usepackage{graphicx}% Include figure files
\usepackage{dcolumn}% Align table columns on decimal point
\usepackage{bm}% bold math
\begin{document}

\preprint{AIP/123-QED}

\title[R. Arun, P. Sabareesan and M. Daniel]{Effect of Transverse Magnetic Field on Dynamics of Current Driven Domain Wall Motion in the Presence of Spin-Hall Effect}% Force line breaks with \\
%\thanks{Footnote to title of article.}
\author{R. Arun}%
\email{arunbdu@gmail.com}
\affiliation{Centre for Nonlinear Dynamics, School of Physics, Bharathidasan University, Tiruchirapalli - 620 024, Tamilnadu, India.%\\This line break forced with \textbackslash\textbackslash
}%
\author{P. Sabareesan}
\email{sabaribard@gmail.com}
 %\homepage{http://www.Second.institution.edu/~Charlie.Author.}
\affiliation{
Centre for Nonlinear Science and Engineering, School of Electrical and Electronics Engineering, SASTRA University, Thanjavur-613 401, Tamilnadu, India.%\\This line break forced% with \\
}%

\author{M. Daniel}
\email{danielcnld@gmail.com}
% \altaffiliation[Also at ]{Physics Department, XYZ University.}%Lines break automatically or can be forced with \\
\affiliation{Centre for Nonlinear Dynamics, School of Physics, Bharathidasan University, Tiruchirapalli - 620 024, Tamilnadu, India.%\\This line break forced with \textbackslash\textbackslash
}%
%\date{\today}% It is always \today, today,
           %  but any date may be explicitly specified
\begin{abstract}
Theoretically, we study the dynamics of a current induced domain wall in the bi-layer structure consists of a ferromagnetic layer and a non-magnetic metal layer with strong spin-orbit coupling in the presence of spin-Hall effect.  The analytical expressions for the velocity and width of the domain wall interms of excitation angle are obtained by solving the Landau-Lifshitz-Gilbert equation with adiabatic, nonadiabatic and spin Hall effect-spin transfer torques using Schryers and Walker's method.  Numerical results show that the occurance of polarity switching in the domain wall is observed only above the threshold current density.  The presence of transverse magnetic field along with spin Hall effect-spin transfer torque enchances the value of the threshold current density, and the corresponding saturated velocity at the threshold current density is also increased.

%Valid PACS numbers may be entered using the \verb+\pacs{#1}+ command.
\end{abstract}

% PACS, the Physics and Astronomy
                             % Classification Scheme.
%Use showkeys class option if keyword
                              %display desired
\maketitle

%%%%%%%%%%%%%%%%%%%%%%%%%%%%%%%%%%%%%%%%%%%%%%%%%%%%%%%%%%%%
\section{Introduction}
The manipulation of a domain wall in a ferromagnetic nanostructure by magnetic field and current play an important role for many important technological applications including logic device\cite{Allwood} and information storage\cite{Parkin}. The field induced domain wall motion can be explained by the reduction in Zeeman energy\cite{Wang}, whereas the current induced domain wall motion is accomplished by the adiabatic and nonadiabatic spin transfer torques\cite{Zhang,Li}. These torques transfer the spin angular momentum from nonequilibrium conduction electrons to local magnetic moments\cite{Zhang}. In general, the motion of the domain wall is in the direction of the applied magnetic field whereas for current it moves in the opposite direction. The regular motion of the domain wall is limited by the Walker breakdown limit and above the limit, the dynamics of the wall changes from regular to oscillatory behavior\cite{Li}.  For the efficient technological applications, the velocity of the domain wall is focused and it is to be enhanced.  The maximum velocity of the domain wall is limited by the Walker breakdown and the increase of this limit is attained by applying the in-plane transverse magnetic field\cite{Glathe, Glathe1, Richter, Kunz, Matthew, Lu}. The present authors also studied the dynamics of current and field induced transverse type Neel domain wall in the presence of transverse magnetic field and observed that the velocity of the domain wall is increased enormously\cite{1st_paper}. 

Recently, a new type of spin-transfer torque due to spin-orbit coupling namely spin Hall effect-spin transfer torque has been identified in a bi-layer system consists of ferromagnetic layer and non-magnetic metal layer with strong spin-orbit coupling\cite{Liu,Liu1,Jiyu,Haazen}. When an in-plane current is passed through the bi-layer system, a perpendicular spin current is produced in the metal layer due to spin-Hall effect and it is injected into the ferromagnetic layer, where the injected spin current exerts the spin Hall effect-spin transfer torque on the magnetic moments. The applications of the spin-Hall effect include storage\cite{Liu}, magnetization switching\cite{Liu,Lee,Finocchio} and the recent works contribute the influence of spin-Hall effect in domain wall dynamics\cite{Haazen,Jisu,Seo,Eduardo,Peng}.  The strength of the spin Hall effect-spin transfer torque is determined by spin-Hall angle which is the ratio between the density of the spin current injected into the ferromagnetic layer and the density of the current passed into the metal layer.  The spin Hall effect-spin transfer torque affects the damping ratio in the weak pinning potential exists in the ferromagnetic layer and results in the reduction of threshold current density for depinning the domain wall\cite{Jisu}. Recently, Haazen et al experimentally proved that the depinning efficiency of the domain wall is increased by spin Hall effect-spin transfer torque \cite{Haazen}. Seo $et~al$ have studied the domain wall motion in Py/Pt bilayer in the presence of spin-Hall effect\cite{Seo} and observed that there is a possibility of domain wall motion which is along the direction of the current and it enhances the velociy of the domain wall enormously and also obtained the polarity switching of the domain wall without oscillatory behavior. 

From the above motivation of increasing the velocity of domain wall, in the present paper the authors study the effect of transverse magnetic field on the dynamcis of the current induced domain wall in the presence of spin-Hall effect analytically and numerically.  The paper is organized as follows: In Section II, the cartoon model of the domain wall and the governing equation of motion for the dynamics of domain wall are presented.  The analytical expression for excitation angle, width and velocity of the domain wall are also obtained. Section III explores the  numerical results of the domain wall dynamics  which are obtained by solving the dynamical equation using Runge-Kutta-4 method and discusses the impact of transverse magnetic field on the current driven domain wall motion in the presence of spin Hall effect-spin transfer torque. Finally, the results are concluded in Section IV. 
\section{Model and dynamics for domain wall motion}
\begin{figure}[!hbtp]
\centering\includegraphics[angle=0,width=0.6\linewidth]{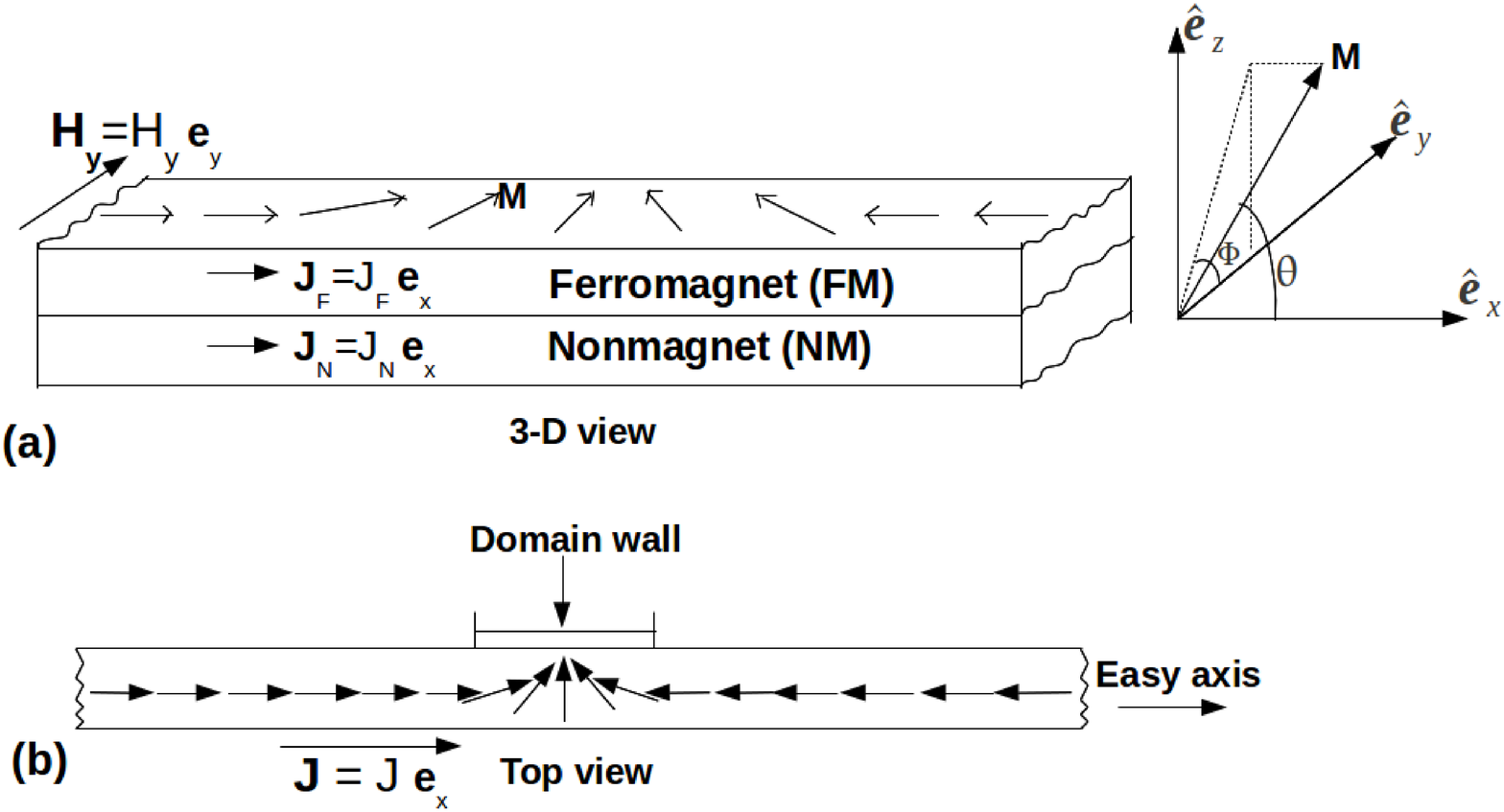}
\caption{(a) A scketch representing a bi-layer model consists of a nonmagnetic metal layer(lower) and a ferromagnetic layer(upper) with an easy axis along x-direction.  The arrow marks indicate the magnetization vector ${\bf M}$.
$J$ is an average applied current density corresponding to the current passed through the bi-layer system along x-direction.
$J_F$ and $J_N$ are the current densities of the current passing through the ferromagnetic and nonmagnetic layers respectively, which are obtained from $J_F$ and $J_N$.  The transverse magnetic field $ H_y$ is applied along the positive y-direction.  (b) Top-view of the bi-layer in the absence of current and transverse magnetic field.}
\label{x}
\end{figure}

A schematic representation of a bi-layer system consists of a ferromagnetic(Py) layer with left and right domains which are seperated by a domain wall and a nonmagnetic(Pt) layer with strong spin-oribit coupling which is responsible for spin-Hall effect has been shown in FIG.1. When a charge current is passed through the bi-layer(Py/Pt) system, the Py layer generates adiabatic and nonadiabatic spin transfer torques and the Pt layer experiences spin-Hall effect and generates spin Hall effect-spin transfer torque.  Assume that the magnetization vector ${\bf M}$ is uniform in y and z directions, $\theta$ is the angle between magnetization vector and positive x-direction which represents the variation of the magnetization along x-direction and $\Phi$ is the angle between the projection of magnetization vector in the yz-plane and positive y-direction which represents the out-of-plane excitation of the magnetization along x-axis. ${\bf H}_y=H_y{\bf \hat e}_y$ is the transverse magnetic field applied along the positive y-direction.  ${\bf J}=J{\bf \hat e}_x$ refers the average applied current density corresponding to the current applied along x-direction in bi-layer system and the current densities in ferromagnetic(${\bf J}_F=J_F{\bf \hat e}_x$) and nonmagnetic(${\bf J}_N=J_N{\bf \hat e}_x$) layers are obtained from $J$.  The dynamics of domain wall in the Py/Pt bi-layer nanostrip is governed by the famous Landau-Lifshitz-Gilbert equation with different spin transfer torques can be expressed as follows.
 %\begin{subequations}
%\label{LLG}
\begin{align}
\frac {\partial {\bf M}}{\partial t}  =  -\gamma {\bf M} \times {\bf H}_{eff} &+ \frac{\alpha }{M_{s}} {\bf M} \times \frac{\partial {\bf M}}{\partial t} -\frac{b}{M_{s}^2} {\bf M} \times \left({\bf M} \times \frac{\partial {\bf M}}{\partial x}\right)\nonumber\\
&- \frac {c}{M_{s}} {\bf M} \times  \frac{\partial {\bf M}}{\partial x}-  \frac{\theta_{SH} d}{M_s} {\bf M\times(M\times {\bf \hat e}_y)}, \label{LLG1}
\end{align}
where,
\begin{subequations}
\label{stt}
\begin{align}
b &= \frac{P J_F}{\mu_B e M_s},\label{stt1}\\
c &= \xi b, \label{stt2}\\
d &= \frac{\gamma \hbar J_N}{2 e M_s t_F}.\label{stt3}
\end{align} 
\end{subequations}
$M_s(=|{\bf M}|)$, $\gamma$ and $\alpha$ refer the saturation magnetization, gyromagnetic ratio and Gilbert damping parameter respectively. $b$ and $c$ are the magnitudes of the adiabatic and nonadiabatic spin transfer torques with the nonadiabaticity parameter($\xi$) respectively. $d$ is magnitude of spin transfer torque due to spin-Hall effect, $\theta_{SH}$ is spin-Hall angle and $P$ is polarization of spin current. $\mu_B$, $e$ and $t_F$ are the Bohr magneton, charge of electron and thickness of the ferromagnetic layer respectively. ${\bf H}_{eff}$ represents the effective field due to different magnetic contributions. In the right hand side of Eq.\eqref{LLG1}, the first term represents the precession of magnetization about the effective field which determines the precessional frequency and conserves the magnetic energy. The second term represents damping of magnetization, which dissipates energy during the precession of magnetization.  The third and fourth terms are adiabatic and non-adiabatic spin transfer torques. In the adiabatic assumption, the polarization of the electrons spin is parallel to the direction of the magnetization, the spatial variation of magnetization in domain wall excerts the torque on the conduction electrons passing through the domain wall and consequently the reaction torque by the conduction electrons is excerted on the domain wall is called adiabatic spin transfer torque. The nonadiabatic spin transfer torque is the reaction torque produced by the mistracking spins between the conduction electron and local magnetization. The last term is the spin Hall effect-spin transfer torque. The effective field contribution present in the bilayer system is given by
\begin{align}
{\bf H}_{eff} = \frac{2A}{M_{s}^2} ~\frac{\partial^2 {\bf M}}{\partial x^2} + \frac{H_{k}}{M_{s}}M_{x} {\bf \hat e}_x + H_y {\bf \hat e}_y - \frac{H_h}{M_s}M_z{\bf \hat e}_z, \label{Heff1}
\end{align}
where, the first term represents the field due to exchange interaction and $A$ is exchange interaction constant . $H_k$ and $H_h$ are the easy axis and hard axis anisotropy fields respectively and $H_y$ is the transverse magnetic field.

In order to understand the dynamics of domain wall, the Eq.\eqref{LLG1} has to be solved, which is a highly nontrivial vector nonlinear evolution equation and difficult to solve in the present form. Hence, Eq.\eqref{LLG1} is transformed into polar coordinates(FIG.1(a)) using the following expressions 
\begin{subequations}
\label{spherical}
\begin{align}
M_{x} &= M_{s} \cos\theta,\label{spherical1}\\
M_{y} &= M_{s} \sin\theta \cos\Phi,\label{spherical2}\\
M_{z} &= M_{s} \sin\theta \sin\Phi.\label{spherical3}
\end{align}
\end{subequations}
The following equations are obtained by substituting Eqs.\eqref{spherical} in Eq.\eqref{LLG1}.
\begin{subequations}
\label{llg}
\begin{align}
\frac{\partial \theta}{\partial t} + \alpha \sin\theta \frac{\partial \Phi}{\partial t} & ~=~  \frac{2\gamma A}{M_s}\left(2 \cos\theta \frac{\partial \theta}{\partial x} \frac{\partial \Phi}{\partial x} + \sin\theta \frac{\partial^2 \Phi}{\partial x^2} \right)-\gamma H_y \sin\Phi\notag\\ &- \frac{\gamma}{2}H_h \sin\theta \sin2\Phi + b \frac{\partial \theta}{\partial x} + c \sin\theta \frac{\partial \Phi}{\partial x}+\theta_{SH} d \cos\theta \cos\Phi, \label{llg1} \\
\alpha\frac{\partial \theta}{\partial t}- \sin\theta \frac{\partial \Phi}{\partial t}& =~ \frac{2\gamma A}{M_{s}}\left[\frac{\partial^2 \theta}{\partial x^2}-\sin \theta \cos\theta\left(\frac{\partial \Phi}{\partial x}\right)^2\right]+\gamma H_y \cos\Phi \cos\theta  \nonumber\\
&-\frac{\gamma}{2}\left[H_k + H_h \sin^2\Phi\right]\sin2\theta  - b \sin\theta \frac{\partial \Phi}{\partial x} + c\frac{\partial \theta}{\partial x}+\theta_{SH} d \sin\Phi. \label{llg2}
\end{align}
\end{subequations}

The above transformed equations describe the dynamics of current induced domain wall motion in a ferromagnetic nanostrip in the presence of transverse magnetic field and spin-Hall effect. These equations are solved to study the dynamical parameters such as excitation angle, velocity and width of the domain wall and the impact of transverse magnetic field on these dynamical parameters in the presence of spin-Hall effect. When the transverse magnetic field is applied along the positive y-direction, it excerts a torque on the magnetic moments in the strip and changes their direction towards positive y-direction.  Consequently, the direction of magnetization in the left and right domains turns symmetrically to the new equilibrium direction towards positive y-direction and correspondingly the angle $\theta$ changes from 0 to $\theta_D$ in left domain and from $\pi$ to $\pi-\theta_D$ in right domain, whereas there is no variation in $\Phi$ and it is zero in the entire strip. The value of $\theta_D$ is given by $\theta_D=\sin^{-1}(H_y/H_k)$\cite{1st_paper}, which forms a constraint $H_y< H_k$.  Otherwise all the magnetic moments of the strip would orient along positive y-direction and it leads to the disappearance of the domain wall, therefore $H_y$ is always taken below $H_k$.
It is assumed that the transverse magnetic field is applied to the static domain wall first along positive y-direction and after the magnetic moments in the strip come to the new equilibrium direction, current is applied along x-direction in the presence of transverse magnetic field.
%   Assume that the transverse magnetic field is applied to the static domain wall along positive y-direction, the magnetization inside the domains change to the new equilibrium direction and then the current is applied in the presence of transverse magnetic field. 
The magnetization inside the domains can be excited in the out-of-plane direction due to spin Hall effect-spin transfer torque and this change in the magnetization direction is neglected for low current densities due to the weak field associated with spin Hall effect-spin transfer torque.

In order to solve the Eqs.\eqref{llg1} and \eqref{llg2}, the trial functions for $\theta$ and $\Phi$ are constructed by Schryer and Walker's method\cite{Schryer} and they are written as\cite{1st_paper}
\begin{align}
\theta(x,t) &= 2\tan^{-1}\left(\frac{a_1+a_2 \exp\left[ \frac{x-X(t)}{W(t)} \right]}{a_2+a_1 \exp\left[ \frac{x-X(t)}{W(t)} \right]} \right),\label{theta1}\\
\Phi(x,t) &= \phi(t) ~ U\left(\frac{x-X(t)}{W(t)} \right),\label{phi1}
\end{align}
where, $a_1=\sqrt{1+\frac{H_y}{H_k}}-\sqrt{1-\frac{H_y}{H_k}},~a_2=\sqrt{1+\frac{H_y}{H_k}}+\sqrt{1-\frac{H_y}{H_k}}$, $X$ is the position of the center of the domain wall, $W$ is the width of the domain wall, and $\phi(=\Phi(X,t)$) is defined as the excitation angle of the domain wall. $U$ is the step function defined as
\begin{align}
U=1~ \mathrm{when}~ &\left(\frac{x-X(t)}{W(t)}\right)<\pi/2 ~\mathrm{and}~U=0~\mathrm{when}~\left(\frac{x-X(t)}{W(t)}\right)>\pi/2\nonumber.
\end{align}

The equations for the excitation angle, velocity and width of the domain wall are obtained by solving Eqs.\eqref{llg1} and \eqref{llg2} for $x=X(t)$ using the trial fucntions Eqs.\eqref{theta1} and \eqref{phi1}[see Ref.12].  The obtained equations are given below
%The equations for the excitation angle, velocity and width of the domain wall are obtained at $x=X(t)$ by substituting the trial fucntions Eqs.\eqref{theta1} and \eqref{phi1} in Eqs.\eqref{llg1} and \eqref{llg2}\cite{1st_paper}. The obtained equations are given below
\begin{eqnarray}
(1+\alpha^2)\frac{d\phi}{dt} = &&-\frac{\alpha\gamma H_h}{2}\sin2\phi - \alpha\gamma H_y \sin\phi + \frac{b(\alpha-\xi_{eff})}{W(t)}\sqrt{\frac{H_k-H_y}{H_k+H_y}},\label{dphi}\\
v=\frac{dX}{dt} = &&\frac{\gamma W(t)}{1+\alpha^2} \sqrt{\frac{H_k+H_y}{H_k-H_y}}\left[H_y \sin\phi+\frac{H_h}{2}\sin2\phi \right]- b\left(\frac{1+\alpha\xi_{eff}}{1+\alpha^2}\right), \label{v}\\
W =&&\frac{W_0}{\sqrt{1+\frac{H_y}{H_k}}} \left[1 + \frac{H_h}{H_k} \sin^2\phi -\frac{H_y}{H_k} \cos\phi+\frac{\alpha \xi b B_{SH}}{\gamma H_k}\cos\phi \right]^{-\frac{1}{2}}. \label{W}
\end{eqnarray}
Where, $W_0 (= \sqrt{2A/H_k M_s})$ is width of the domain wall in the absence of current and transverse magnetic field,
\begin{align}
\xi_{eff} &= \xi \left(1+B_{SH} \sqrt{\frac{H_k+H_y}{H_k-H_y}}W(t)\sin\phi \right),\label{xi_eff}\\
B_{SH} &= \frac{\pi \theta_{SH} J_N}{2\xi t_F P J_F}.\label{Bsh}
\end{align}

Equations \eqref{dphi}, \eqref{v} and \eqref{W} represent the rate of change of excitation angle($\phi$), velocity($v$) and width($W$) of the domain wall driven by current in the presence of transverse magnetic field along with spin-Hall effect respectively.  Further, Eqs.\eqref{v} and \eqref{W} are depend on $\phi$, hence it is essential to solve the Eq.\eqref{dphi} to find velocity and width of the domain wall. %Initally the value of $\phi(0)$ is 0 in the presence of current without transverse magnetic field and the corresponding the width W(0) is $W_0$, but the analytical expression (\eqref{W}) shows that the initial width $W(0)=W_0/\sqrt{1 + ({\alpha \xi bB_{SH}}/{\gamma H_k}) }\approx W_0$. This small discripancy in the initial width is due to the exclusion of the change in magnetization inside the domains due to spin-Hall effect for low current densities. 

\section{NUMERICAL RESULTS AND DISCUSSION}
In the previous section, the analytical expression for width and velocity of the domain wall have been derived, but the excitation angle is not explicitly derived from Eq.\eqref{dphi} becasue it is a nontrivial nonlinear evolution equation and very difficult to solve analytically.  Hence, in this section Eq.\eqref{dphi} is numerically computed using Runge-Kutta-4 method for the initial condition $\phi(0)=0$.  For numerical calculation, the geometry and the material parameters are given as follows: bi-layer struture made of Py and Pt with dimensions 2 micron(length), 80 nm(width), 4 nm(thickness of Py) and 3 nm(thickness of Pt) and the material parameters\cite{Seo,Sabareesan} are given as $M_s=8.0\times 10^5 A/m,~A=1.3\times10^{-11}~J/m,~P=0.7,~\alpha=0.02$, $\xi=0.01$, $H_k = 3.979\times10^3 ~A/m$ for Py and  $\theta_{SH}=\pm 0.1$ for Pt.  $J_F$ and $J_N$ are computed from average applied current density $J$ using the circuit model as follows\cite{Seo}: $J_F=J(t_F+t_N)\sigma_F/(t_F\sigma_F+t_N\sigma_N)$ and $J_N=J(t_F+t_N)\sigma_N/(t_F\sigma_F+t_N\sigma_N)$, where $\sigma_F$($\sigma_N$) is the conductivity of the ferromagnetic(nonmagnetic) layer and $t_N$ is the thickness of the nonmagnetic layer. Let as consider the conductivity of both the layers are same (i.e. $\sigma_{Py}=\sigma_{Pt}=6.5~(\mu \Omega m)^{-1}$) and it leads to $J_F = J_N=J$. Here we discuss the numerical results of the excitation angle in the absence and presence of transverse magnetic field, and the effect of transverse magnetic field on threshold current density and saturated velocity.

The excitation angle $\phi$ in the absence of transverse magnetic field for different current densities $J=\pm 1.0\times 10^{12}~A/m^{2}$, $J=\pm 1.009\times 10^{12}~A/m^{2}$ and $J=\pm 1.01\times 10^{12}~A/m^{2}$  with spin-Hall angles $\theta_{SH}=\pm 0.1$ have been plotted in FIGs.2.  The excitation angle starts from 0$^\circ$ and it reaches the saturated excitation angle($\phi_s$) at $\pm$ 3.05$^\circ$  and $\pm$ 5.18$^\circ$ for $J=\pm 1.00\times 10^{12}~A/m^{2},\theta_{SH}=\mp 0.1$ and  $J=\pm 1.009\times 10^{12}~A/m^{2},\theta_{SH}=\mp 0.1$ respectively as shown in FIG.2(a). The upward ($J>0,\theta_{SH}=-0.1$) or downward ($J<0,\theta_{SH}=+0.1$) excitation of the domain wall can be controlled by the sign of the spin-Hall angle and direction of the current, because of the interaction between nonadiabatic spin Hall effect-spin transfer torques. For $J < J=\pm 1.01\times 10^{12}~A/m^{2}$, the excitation angle of the domain wall increases with respect to time and after few nanoseconds it reaches its saturation and there is no switching in the polarity of the domain wall occurs, which has been shown in the inset figure of Fig.2(a) by plotting normalized magnetization along y-direction at the centre of the domain wall ($\left\{\frac{M_y}{M_s}\right\}_{x=X}$) against time.  If we increase the current density from $J=\pm 1.009\times 10^{12}~A/m^{2}$ to $J=\pm 1.01\times 10^{12}~A/m^{2}$, the exciation angle slowly increases from 0$^\circ$ and maintains constant value upto $\sim$300 ns and after that there is a drastic change in the excitation angle from $\sim 0^\circ$ to $\sim 180 ^\circ$ or $-180^\circ$ with respect to the sign of spin-Hall angle and direction of the current(see FIG.2(b)).  

In this case, the polarity switching occurs and the polarity of the domain wall gets reversed, which can be observed by plotting the normalized magnetization $\left\{\frac{M_y}{M_s}\right\}_{x=X}$ along y-direction against time(see inset figure (i) of FIG.2(b)).  The $\left\{\frac{M_y}{M_s}\right\}_{x=X}$ is initially at one stable state (+1) and after 300 ns it switches to  another stable state (-1) when the current density $J$ is increased above $\pm 1.009\times 10^{12}~A/m^{2}$ for $\theta_{SH}$ is $\mp$0.1 respectively. Hence the value of the current density $J$ for polarity switching is greater than $\pm 1.009\times 10^{12}~A/m^{2}$ which can be referred as the threshold current density($J_p$).  The polarity switching of the domain wall occurs only when the spin Hall effect-spin transfer torque dominates the nonadiabatic spin transfer torque while both torques are in the opposite directions.  However, in the case of $J<0$ or $>0$ and $\theta_{SH}=$-0.1 or +0.1 respectively, the polarity switching cannot be occured because the directions of spin Hall effect-spin transfer torque and nonadiabatic spin transfer torque are act in same direction.

As the excitation angle $\phi$ saturates to $\phi_s$ with respect to time, the corresponding width($W$) and velocity of the domain wall($v$) also saturate to $W_s(=W(\infty))$ and $v_s(=v(\infty))$ respectively. The corresponding saturated velocity $v_s$ with respect to the current density $J$ for $\theta_{SH}=\pm 0.1$ is shown in the inset figure (ii) of FIG.2(b). It shows that, when the sign of current density and spin-Hall angle are same, the direction of saturated velocity of the domain wall is opposite to the direction of current density, which means that the domain wall moves along the electron flow direction. For this case, the spin-Hall effect and nonadiabatic spin transfer torques act in the same direction, whereas when the sign of $J$ and $\theta_{SH}$ are opposite and the current density $J$ is increased towards the threshold current density, the magnitude of the saturated velocity reaches the maximum value at $J=J_p$ and its direction changes into the direction of current density which implies that the motion of domain wall is in the opposite direction of the electron flow. The saturated velocity at threshold current density($\{v_s\}_{J=J_p}$) is $\sim\pm700 ~m/s$ for the $\theta_{SH}=\mp0.1$ respectively. Above the threshold current density, the magnitude of the saturated velocity is suddenly reduced from $\pm700 ~m/s$ and its direction turns back to the direction of current density due to the polarity switching. 

\begin{figure}[htb]
\centering\includegraphics[angle=0,width=0.5\linewidth]{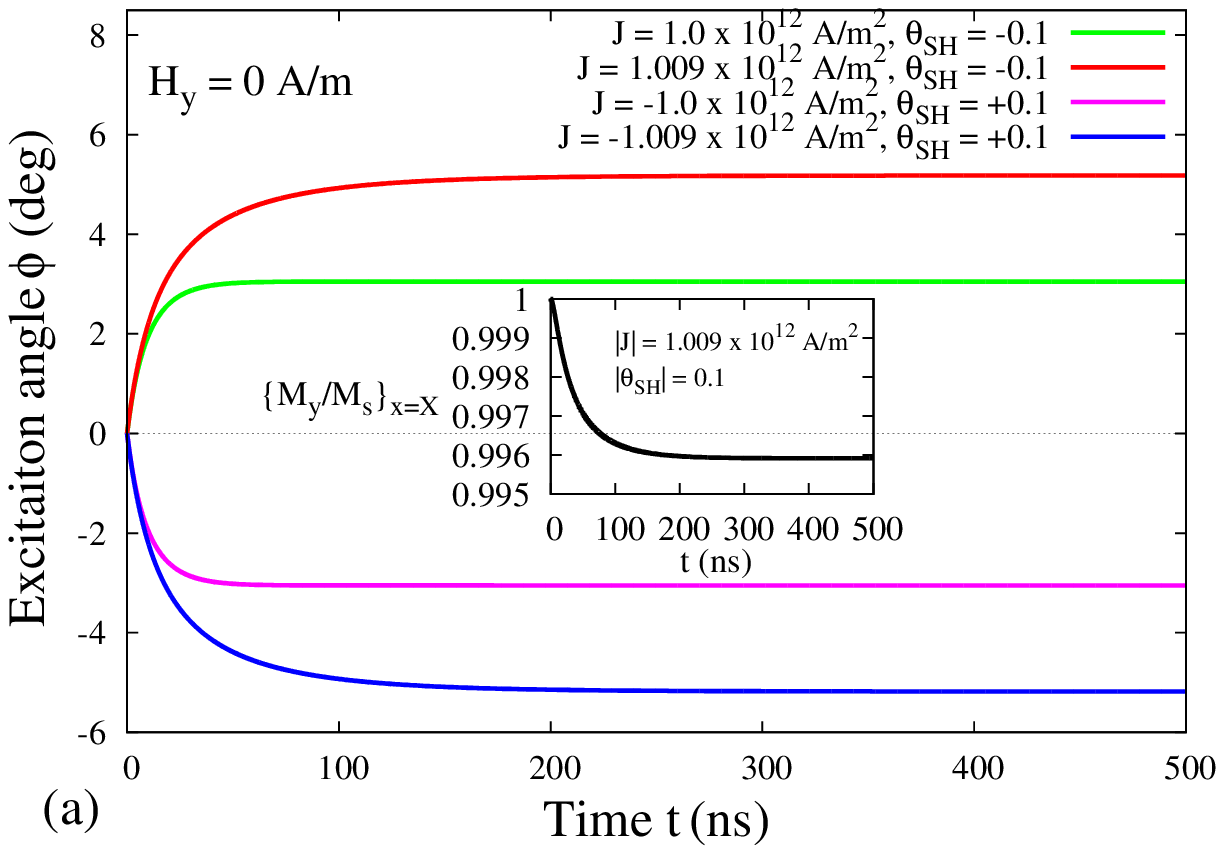}\includegraphics[angle=0,width=0.5\linewidth]{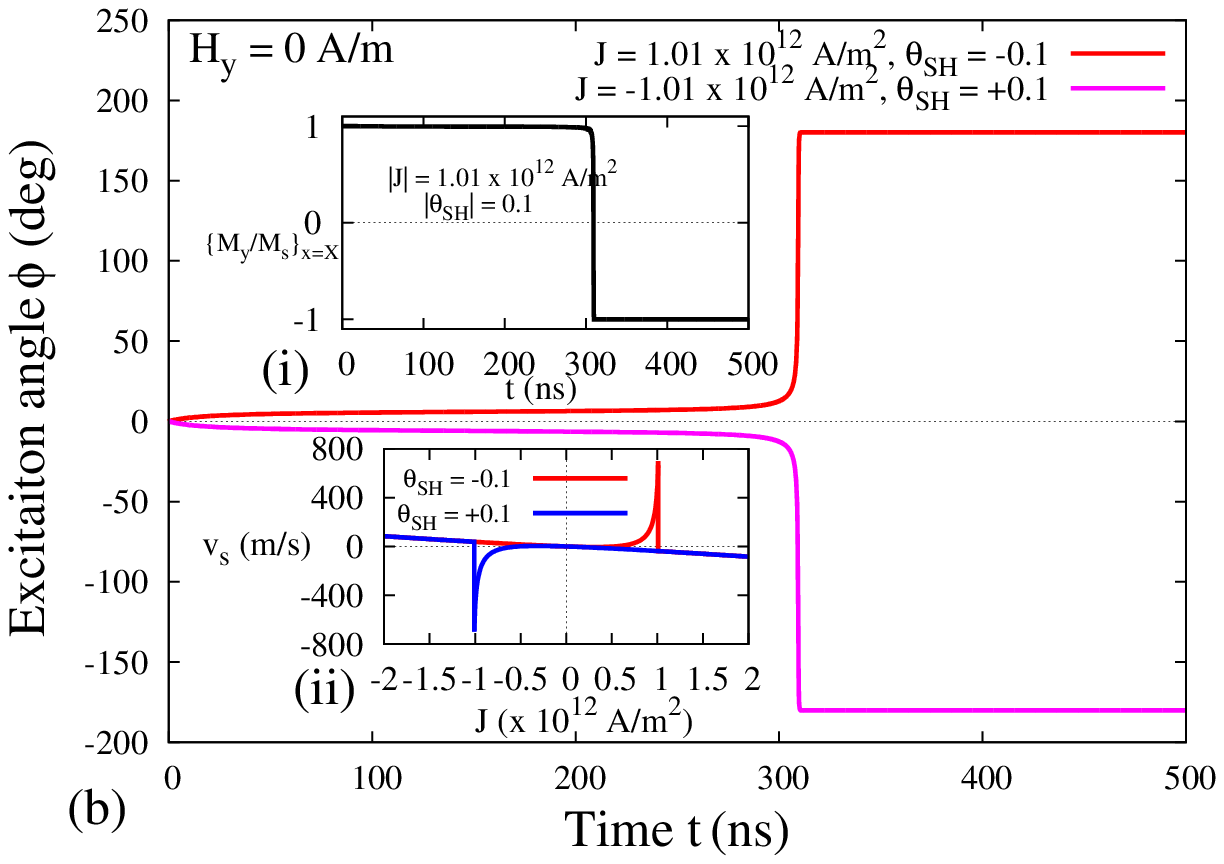}
\caption{(a) The excitation angle $\phi$ with respect to time $t$ for the differnt current densities $J=\pm1.000\times 10^{12}~A/m^{2}$, $\pm1.009\times 10^{12}~A/m^{2}$ for $\theta_{SH}=\mp0.1$ respectively. The inset figure of (a) shows the corresponding variation in the y-component of the normalizedf magnetization at the center of the domain wall $\left\{\frac{M_y}{M_s}\right\}_{x=X}$ with time.  (b) The time variation of the excitation angle for the current density $J=\pm1.01\times 10^{12}~A/m^{2}$ for $\theta_{SH}=\mp0.1$ and the corresponding variation of $\left\{\frac{M_y}{M_s}\right\}_{x=X}$ (inset figure (i) of FIG.2(b)).  The saturated velocity of domain wall against current density $J$ for the spin-Hall angles $\pm$0.1. in the absence of transverse magnetic field(inset figure (ii) of FIG.2(b)).}
\end{figure}

\begin{figure}[htb]
\centering\includegraphics[angle=0,width=0.8\linewidth]{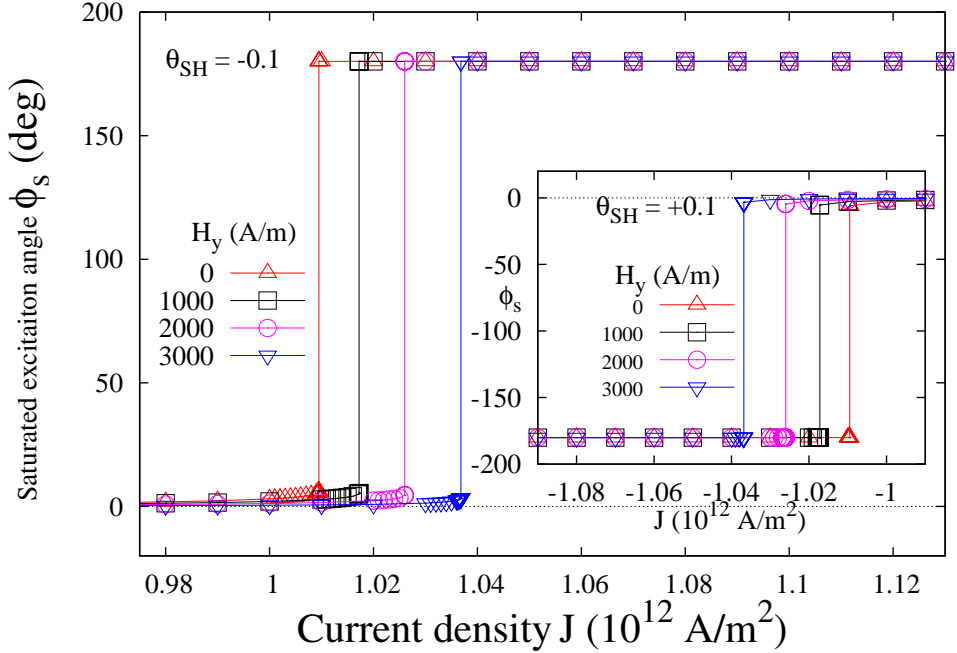}
\caption{The variation of saturated excitation angle $\phi_s$ with respect to current density $J$ in the presence of transverse magnetic fields $H_y=$ 0 $A/m$, 1000 $A/m$, 2000 $A/m$ and 3000 $A/m$ for $\theta_{SH}=$ -0.1 and $\theta_{SH}=$ +0.1(inset figure).}
\end{figure}

So far we discussed the behaviour of the excitation angle with respect to different current densities and spin-Hall angles in the absence of transverse magnetic field.  It leads to the polarity switching of domain wall above the threshold current density $J_p$ and the maximum saturated velocity $\sim700 ~m/s$ at $J=J_p$ in the absence of transverse magnetic field. Here we study the effect of a transverse magnetic field on current induced domain wall motion in the presence of spin-Hall effect in order to enhance the velocity of the domain wall. For that, first we study the behaviour of the saturated excitation angle by varying the transverse magnetic field. The saturated excitation angle($\phi_s$) against the current density $J$ for different transverse magnetic fields $H_y$ = 0 $A/m$, 1000 $A/m$, 2000 $A/m$ and 3000 $A/m$ with $\theta_{SH}=$ -0.1 is plotted in FIG.3 and in the inset figure of FIG.3 with $\theta_{SH}=$ 0.1. There is a slight increment in the magnitude of threshold current $|J_p|$ by varying the strength of the transverse magnetic field, is observed from FIGs.3.  The variation in $J_p$ is due to a torque created by the transverse magnetic field which acts in the direction of nonadiabatic spin transfer torque. This torque is added along with the nonadiabatic spin transfer torque and the spin Hall effect-spin transfer torque is dominated, hence the strength of the spin Hall effect-spin transfer torque has to be enhanced by increasing the current density to attain the polarity switching.  Therefore the threshold current density is slightly increased by transverse magnetic field and the values of the threshold current densities corresponding to the transverse magnetic fields (0, 1000, 2000 and 3000 $A/m$) are listed in Table.1.

\begin{figure}[htb]
\centering\includegraphics[angle=0,width=0.5\linewidth]{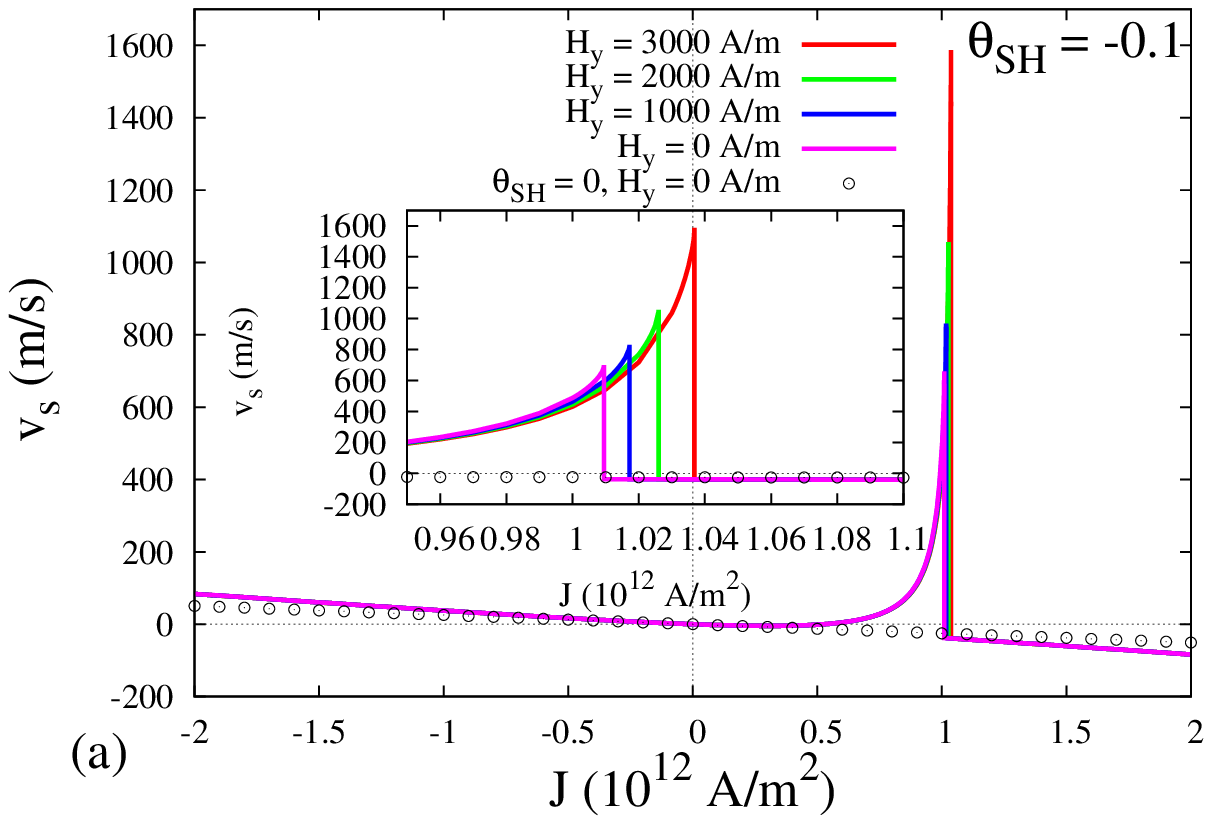}\includegraphics[angle=0,width=0.5\linewidth]{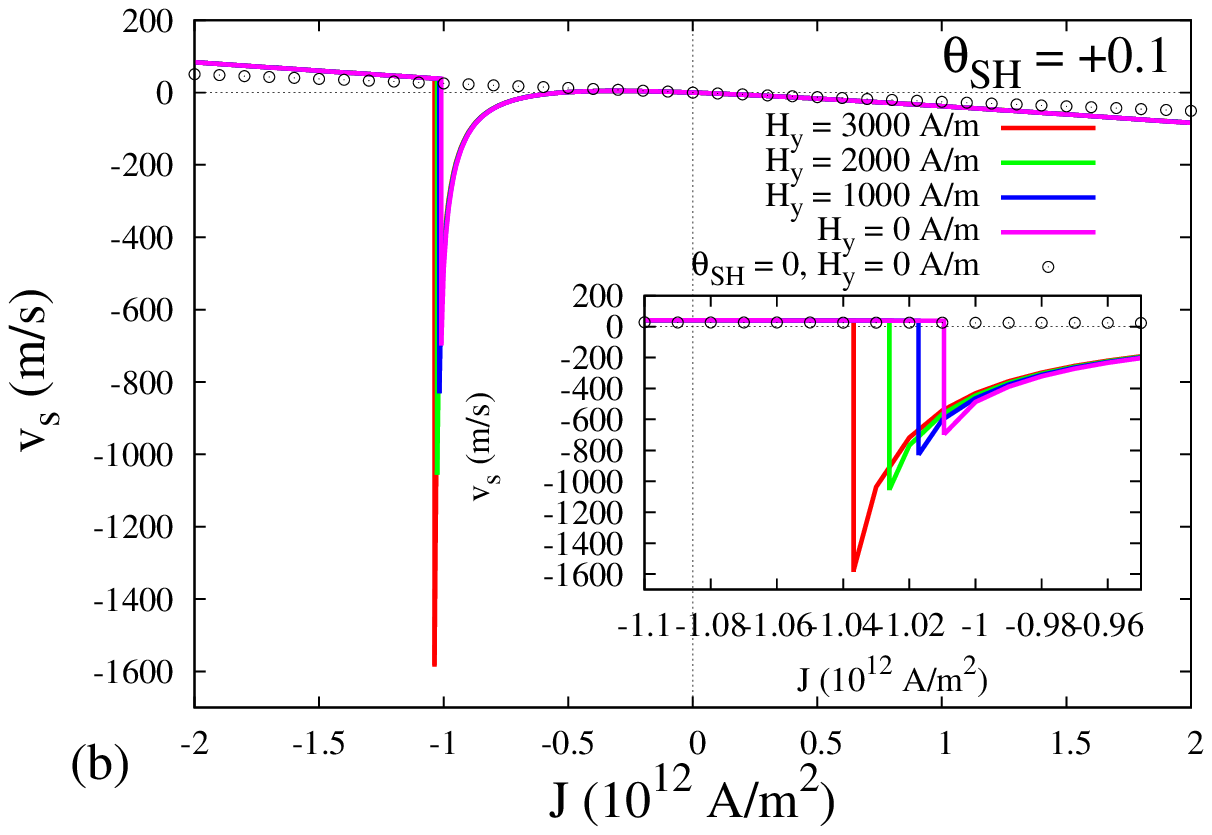}
\caption{The saturated velocity of domain wall $v_s$ with respect to current density $J$ corresponding to (a) $\theta_{SH}=$-0.1 and (b) $\theta_{SH}=$+0.1 for the different transverse magnetic fields $H_y=$ 0 $A/m$, 1000 $A/m$, 2000 $A/m$, 3000 $A/m$.  The inset figures are the zoomed version of the corresponding plots in figures (a) and (b) respectively.}
\end{figure}

In order to study the effect of transverse magnetic field on saturated velocity $v_s$ of the current induced domain wall in the presence of spin-Hall effect, we plot $v_s$ against $J$ from -2.0 $\times$ 10$^{12}$ A/m$^2$ to 2.0 $\times$ 10$^{12}$ A/m$^2$ for $H_y=$0 $A/m$, 1000 $A/m$, 2000 $A/m$, 3000 $A/m$  and $\theta_{SH}=0,\pm0.1$ in FIGs.4(a) and 4(b) respectively. In both the figures, the plots corresponding to $H_y=0$, plotted with open circle and pink color, exactly coincide with the Soo-Man Seo et.al results\cite{Seo}. The results in FIGs.4 show that the saturated velocity of the domain wall gets maximum at the threshold current densities corresponding to $H_y=$ 0 $A/m$, 1000 $A/m$, 2000 $A/m$ and 3000 $A/m$ due to spin-Hall effect and $\{v_s\}_{J=J_p}$ increases considerably with the strength of the transverse magnetic field as shown in the zoomed inset figures of FIGs.4(a) and 4(b).  The observed value of $\{v_s\}_{J=J_p}$ for different transverse magnetic fields are listed in Table.1.

\begin{table}[h]
\begin{tabular}{|c|c|c|c|c|}
\hline
& \multicolumn{2}{|c|}{$\theta_{SH}=-0.1$}& \multicolumn{2}{|c|}{$\theta_{SH}=+0.1$}\\
\cline{2-5}
$H_y$  & $J_p$ & $\{v_s\}_{J=J_p}$ & $J_p$ & $\{v_s\}_{J=J_p}$\\

($A/m$)  & ($\times10^{12}~A/m^2$) & ($m/s$) & ($\times10^{12}~A/m^2$) & ($m/s$)\\
\hline
0    & ~1.009496~  & ~699.05~  & ~-1.009496~  & ~-699.05~ \\
1000 & ~1.017223~  & ~830.65~  & ~-1.017223~  & ~-830.65~ \\
2000 & ~1.026012~  & ~1056.64~ & ~-1.026012~  & ~-1056.64~ \\
3000 & ~1.036829~  & ~1586.72~ & ~-1.036829~  & ~-1586.72~ \\
\hline
\end{tabular}
%\aligning
\caption{The threshold current density $J_p$ and the saturated velocity at threshold current density $\{v_s\}_{J=J_p}$ corresponding to the transverse magnetic fields $H_y=$ 0 $A/m$, 1000 $A/m$, 2000 $A/m$ and 3000 $A/m$ for the spin-Hall angles $\theta_{SH}=$ -0.1 and +0.1.}
\end{table}

The threshold current density $J_p$ and the saturated velocity at threshold current density $\{v_s\}_{J=J_p}$ are plotted  respectively in FIG.5(a) and 5(b) with respect to transverse magnetic field for $\theta_{SH}=\pm$0.1.  Figure 5(a) shows that the threshold current density is slightly increased from 1.009 $A/m$ to 1.037 $A/m$ by varying the transverse magnetic field from 0 $A/m$ to 3000 $A/m$ respectively. Because of the increase in threshold current density, $\{v_s\}_{J=J_p}$ is also increased (see Fig.5(b)).  This can be verified analytically from Eq.\eqref{stt3} as the strength of the spin Hall effect-spin transfer torque is directly proportional to the current density  $J$, which implies that the threshold current density is increased and the corresponding strength of the spin Hall effect-spin transfer torque is also increased. Hence the saturated velocity of the domain wall at the threshold current density is increased.
\begin{figure}[htb]
\centering\includegraphics[angle=0,width=0.5\linewidth]{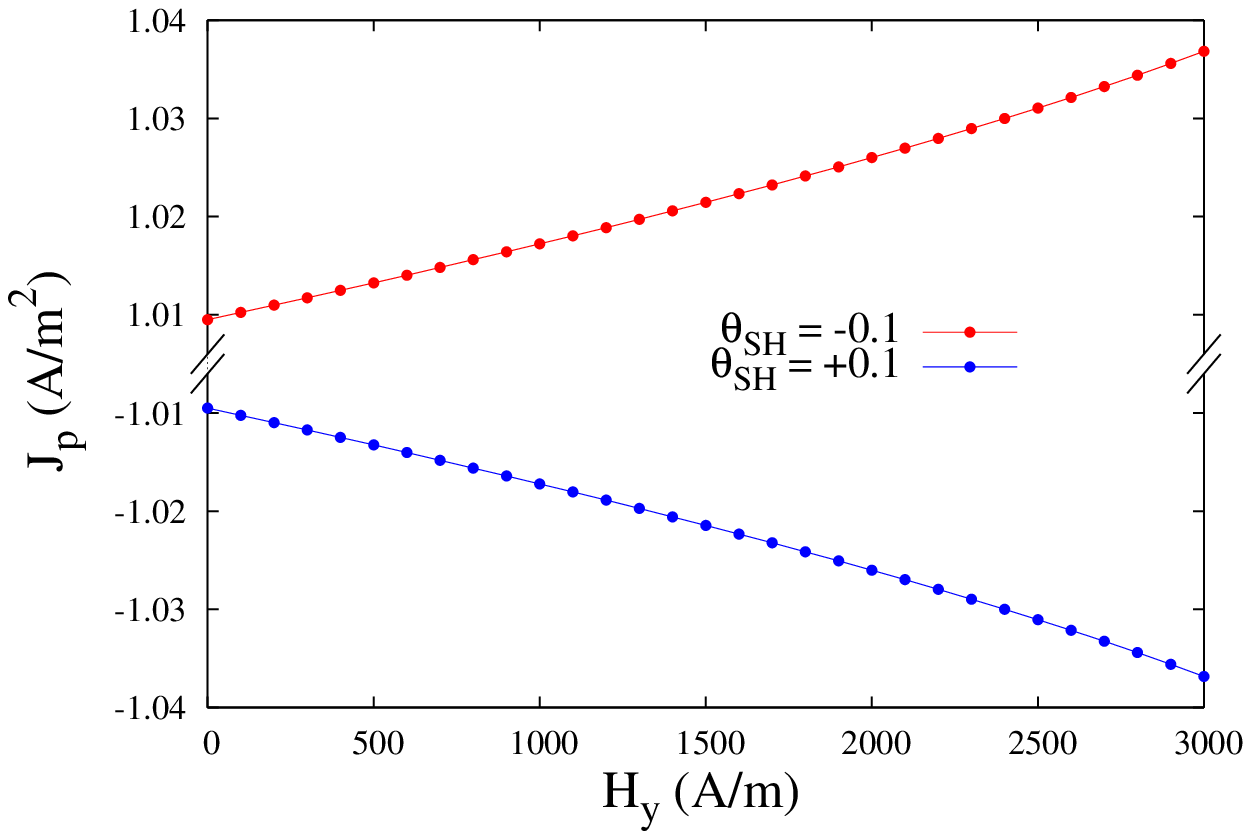}~\includegraphics[angle=0,width=0.5\linewidth]{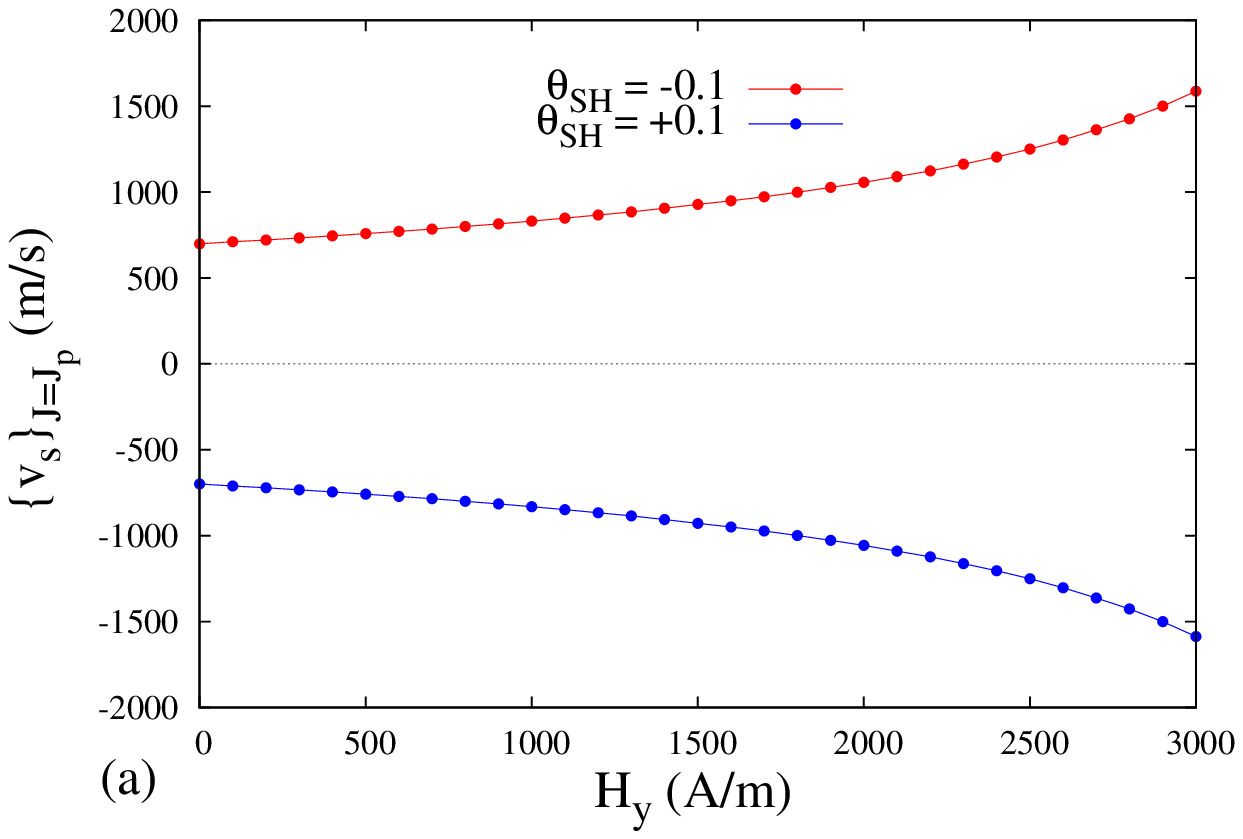}
\caption{(a) The threshold current density $J_p$ and (b) saturated velocity at the threshold current density $\{v_s\}_{J=J_p}$ with respect to the transverse magnetic field $H_y$ corresponding to the spin-Hall angles $\pm$0.1.}
\end{figure}

\section{Conclusions}
In the present paper, the authors study the current driven domain wall dynamics in a Py/Pt bi-layer structure in the presence of spin-Hall effect along with transverse magnetic field. The dynamics is governed by Landau-Lifshitz-Gilbert equation with adiabatic, nonadiabatic spin transfer torques and spin Hall effect-spin transfer torque, which was solved by Schryer and Walker's method. From the results, the analytical expression for the velocity and width of the domain wall are expressed in terms of the excitation angle. The numerical results show that the polarity switching of the domain wall occurs only when the current density is above the threshold, which is confirmed by the increase of saturated excitation angle from $\sim$0$^\circ$ to $\sim\pm$180$^\circ$ corresponding to the spin-Hall angles $\mp$0.1. The effect of the transverse magnetic field shows that the threshold current density increases from 1.009 $A/m^2$(-1.009 $A/m^2$) to 1.037 $A/m^2$(-1.037 $A/m^2$) when the transverse magnetic field is increased from 0 $A/m$ to 3000 $A/m$ corresponding to the spin-Hall angle -0.1(+0.1).  Consequently, the corresponding saturated velocity at the threshold current density also enhances from 699.05 $m/s$(-699.05 $m/s$) to 1586.72 $m/s$(-1586.72 $m/s$) due to the increase in transverse magnetic field from 0 $A/m$ to 3000 $A/m$ corresponding to the spin-Hall angle -0.1(+0.1). 
 
In conclusion, the polarity switching of the domain wall occurs only above the threshold current density, the increase in transverse magnetic field increases the threshold current density and the corresponding saturated velocity at the threshold current density is also increased. From the above results, we observed that the in-plane transverse magnetic field plays an important role in increasing the velocity of the current induced domain wall dynamics in the bi-layer structure with the spin-Hall effect and it may be useful in developing high performance applications using domain wall motion. 

\section{Acknowledgement}
The work of M.D and R.A forms part of a major DST project. P.S wishes to thank DST
Fast track scheme for providing financial support.

\end{document}